\documentclass[journal,final,letterpaper,twoside,twocolumn]{IEEEtran}
\usepackage[noadjust]{cite}      
\usepackage{graphicx}  
\usepackage{subfigure} 

\newtheorem{thm}{\bf Theorem}

\begin{document}
\pagestyle{empty} \thispagestyle{empty}
\title{Minimal Network Coding for Multicast}
\author{\authorblockN{Kapil Bhattad\authorrefmark{1},
Niranjan Ratnakar\authorrefmark{2}, Ralf Koetter\authorrefmark{2},
and Krishna R. Narayanan\authorrefmark{1}}
\\\authorblockA{\authorrefmark{1}Texas A \& M University, College Station,
TX. Email: kbhattad,krn@ee.tamu.edu}
\\\authorblockA{\authorrefmark{2}University of Illinois, Urbana Champaign, IL. Email: ratnakar,koetter@uiuc.edu}
\thanks {This work was supported by National Science Foundation
under grant CCR 0093020.}} \maketitle \thispagestyle{empty}
\begin{abstract}

We give an information flow interpretation for multicasting using
network coding. This generalizes the fluid model used to represent
flows to a single receiver. Using the generalized model, we
present a decentralized algorithm to minimize the number of
packets that undergo network coding. We also propose a
decentralized algorithm to construct capacity achieving multicast
codes when the processing at some nodes is restricted to routing.
The proposed algorithms can be coupled with existing decentralized
schemes to achieve minimum cost muticast.

\end{abstract}


\section{Introduction}

\label{sec:introduction}

In their seminal work,  Ahlswede {\em et al} \cite{ahlswede}
showed that if the nodes in the network are allowed to perform
network coding rather than just routing then the max flow min cut
bound on the multicast capacity is achievable. Li {\em et al}
\cite{li} showed that linear codes are sufficient to achieve the
multicast capacity. Since then several techniques have been
proposed to design codes that achieve the multicast capacity.
Among them, the idea of random network coding seems very
promising. Ho {\em et al} \cite{ho,ho2} propose a scheme in which
data is collected in the form of packets of, say, length $n$.
These packets are then treated as elements of a finite field of
size $q = 2^n$ (assuming that the data is in bits) and they show
that if the messages on the outgoing edges of every node are set
to be a random linear combination of the messages received along
the incoming edges over a finite field of size $q$ then the
probability that the resulting code is not a valid multicast code
is $O(1/q)$. (We call a multicast  {\em valid} if the destination
nodes can decode the data.) Therefore a valid multicast code can
be designed with very high probability by random coding over a
large field.

Random network coding by itself could be inefficient in terms of
network resources. Since the scheme is completely distributed and
there is no communication between the nodes, each node sends
messages on all its outgoing edges in the process using up all the
available bandwidth. But, this problem can be solved. In
\cite{lun,lun2} Lun {\em et al} proposed a distributed algorithm
which can be used, for example, to find a sub network that
minimizes link usage costs while having the same multicast
capacity as the given network. Random network coding can be
employed on this sub network to achieve the multicast capacity.

In general, minimal cost network coding solutions are of practical
interest.  The cost to be minimized may depend on the network and
application at hand.  For example, if a router that employs
network coding is expensive we will want to minimize the number of
nodes that perform network coding. In optical networks the
operation of computing linear combination of inputs may require
conversion from optical signals to electrical signals which is
expensive and hence we may want to minimize the number of packets
that undergo network coding. Random network coding as such would
result in
 schemes where every node performs network coding. In this paper, we
will address the problem of minimal cost network coding where the
cost is the number of packets that need to be network coded. We
also consider the problem of finding minimum cost solutions when
some of the nodes are restricted to perform only routing. The
multicast capacity for a special case of this problem when all the
nodes only route has been studied in \cite {cannons}. We will
refer to nodes employing network coding by network coding nodes
and nodes restricted to routing by routing nodes.

In \cite{lun}, the authors consider costs such as bandwidth and
delay and investigate minimum cost multicast. However, the results
in \cite{lun} cannot be directly used to solve the problems
considered in this paper because the fluid model used to represent
flows to individual receivers cannot be used when some of the
nodes are restricted to routing.  It is also not possible to
differentiate between the operations of network coding and routing
at a node by only looking at the input and output flows of that
node. The main contribution of this paper is to give a new
information flow based interpretation for the multicast flow and
use this model to set up optimization problems that can be solved
in a distributed manner.

The optimization problem formulated in this paper
has a complexity that grows exponentially with the number of
receivers but in many applications like video conferencing the
number of receivers is quite small and hence these algorithms can
be of practical use.

In section \ref{sec:notation} we give the notation used in the
paper. We present the new information flow model in section
\ref{sec:infoflow}. In section \ref {sec:optimize} we set up the
optimization problems and finally conclude in section
\ref{sec:conclusion}.

\section{Notation}

\label{sec:notation} We represent a network by a directed graph
${\cal G}=({\cal V}, {\cal E})$, where ${\cal V}$ is the set of
vertices (nodes) and ${\cal E}$ is the set of edges (links). The
capacity of edge $e \in {\cal E}$ is given by $C(e)$. For each
node $v \in {\cal V}$ we define sets $E_{in}(v)$ and $E_{out}(v)$
as the set of all edges that come into $v$ and that go out of $v$
respectively.

We consider a multicast problem with one source $S \in {\cal V}$
and $K$ receivers in the set $D \subset {\cal V}$. We assume that
$D = \{1,2,\cdots,K\}$. For convenience we define two sets ${\cal
P}$ and ${\cal Q}$ where ${\cal P}$ is the power set of $D$
(neglecting the empty set) and ${\cal Q}$ is a set containing all
collections of two or more disjoint sets in ${\cal P}$. For
example, when $K = 3$, ${\cal P} = \{\{1\},$ $ \{2\}, \{3\},
\{1,2\}, \{1,3\}, \{2,3\}, \{1,2,3\}\}$ and ${\cal Q} = \{\
\{\{1\},\{2\}\},\ \{\{1\},\{3\}\},\ \{\{2\},\{3\}\},\
\{\{1\},\{2\},\{3\}\},\ \{\{1\},\\\{2,3\}\},\ \{\{2\},\{1,3\}\}, \
\{\{3\},\{1,2\}\}\ \}$. We fix the ordering in ${\cal P}$ and
${\cal Q}$ and represent the $i$-th element in ${\cal P}$ and the
$j$-th element in ${\cal Q}$ by $P_i$ and $Q_j$ respectively.

With each edge $e \in {\cal E}$ and a coding scheme we associate a
$2^K - 1$ length information flow vector $X_e$ where the $i$-th
element denoted by $x_e(P_i)$ represents the amount of information
common to and only common to receivers in the set $ P_i$ that
flows through the edge $e$. We define $I_k(X_e)= \sum_{i : k \in
P_i} x_e(P_i)$ as the amount of flow along edge $e$ in the flow
decomposition of receiver $k$. The definitions will be made
precise in Section \ref{sec:infoflow}.






It is sometimes convenient to assume that the edges capacities and
the flow vectors are integers. This assumption is justified since
we can always consider the network over multiple time instances.


\section{Information Flow}
\label{sec:infoflow}

In a multicast setup, any multicast solution can be decomposed
into flows to individual receivers \cite{ahlswede}, \cite{li}. The
flows to different receivers could overlap. Overlapping flows
indicates that the data sent along the overlapping part of the
flow has to be eventually conveyed to all the receivers whose
flows overlap.

The  main idea here is to partition the flows to the individual
receivers as components of the form $x_e(P_i)$. We formally do
this in the rest of the section. We define $x_e(\{k_1, k_2,\cdots,
k_j\})$ for an edge $e$ as the amount of overlap in the flows
along $e$ from the source node to the receiver nodes ${k_1},
{k_2}, \cdots, {k_j}$ and that does not overlap with any other
flow for any other receiver. To identify the overlapping flows,
consider a network obtained by expanding the original network by
replacing each edge $e$ by parallel edges,
$e_1',\cdots,e_{C(e)}'$, of unit capacity (assuming edge
capacities are integers). The expanded network also supports the
same rate (h, also assumed to be an integer) as the original
network and hence $h$ edge disjoint paths from source to receiver
$k$ for each $k$ can be found \cite{ahlswede}, \cite{li}. The
paths to the different receivers could have overlapping edges. For
an edge $e'$ in the expanded network, let $P_i$ be the set of all
receivers that have edge $e'$ in one of their paths. The element
$x_{e'}(P)$ in the information flow vector for $e'$ is then $1$
for $P = P_i$ and $0$ otherwise. If no paths pass through $e'$ its
information flow vector is zero. The information flow vector for
the edge $e$ in the original network is the sum of the information
flow vectors of the parallel edges $e_1',\cdots,e_{C(e)}'$.

To keep the notation brief we also use $x_e(k_1, k_2,\cdots, k_j)$
with $k_1 < k_2 < \cdots< k_j$ to represent $x_e(\{k_1,
k_2,\cdots, k_j\})$. 
It is east to see that the flow to receiver $k$ along edge $e$ is
given by $\sum_{i : k \in P_i} x_e(P_i)$. Since this is a function
of $X_e$ for each $k$, we represent it by $I_k(X_e)$. We show the
flow vector and information flow vector for some multicast
networks in Example 1.

\begin{figure}
\includegraphics{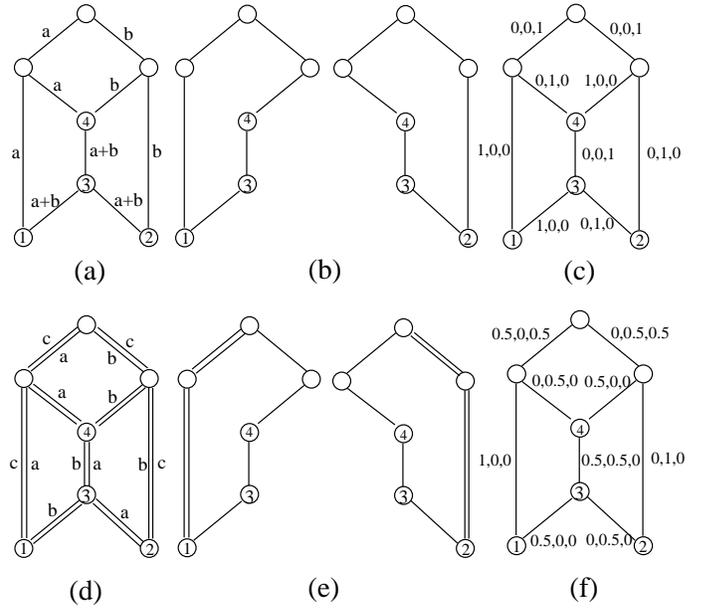}
\caption{Multicast Flows} \label{fig:flow}
\end{figure}

{\bf Example 1}: Consider the network shown in Fig.
\ref{fig:flow}a. A code that achieves the multicast capacity is
shown in Fig. \ref{fig:flow}a. The flows to the two receivers are
shown in Fig. \ref{fig:flow}b. In Fig. \ref{fig:flow}c the
information flow vector for each edge is shown. The information
flow vector is (1,0,0) when the edge carries data at unit rate
only for receiver 1, is (0,1,0) when the edge carries data at unit
rate for receiver 2 and (0,0,1) when the edge carries data at unit
rate meant for both the receivers. In Fig. \ref{fig:flow}d we show
a code over two time instances that achieves the routing capacity
of the network \cite{cannons} and in Fig. \ref{fig:flow}e we show
the corresponding flows. We note that the edge between node 4 and
node 3 has flows for both the receivers but they are not
overlapping flows. It is easy to verify in Fig. \ref{fig:flow}b,
\ref{fig:flow}c, \ref{fig:flow}e, and \ref{fig:flow}f that
$I_1(X_e)$ and $I_2(X_e)$ gives the amount of flow along edge e to
receiver 1 and 2 respectively.

The amount of data flowing along an edge $e$ is the sum of the
elements of $X_e$. Since each edge has a capacity constraint, we
have the following constraint on $X_e$.
\begin{equation}
\label{eqn:edgeconst} \mbox{sum}(X_e) = {\bf b^T}X_e \leq C(e) \ \
\forall\ e \in {\cal E}
\end{equation}
where ${\bf b}$ is the all one vector of length $2^K-1$. We denote
the constraint in (\ref{eqn:edgeconst}) as the {\em edge
constraint}.

\begin{thm}
\label{lemma:flowconst} In any multicast network that supports a
rate $h$, we can find $X_e$ for every edge $e \in {\cal E}$
satisfying the edge constraint such that
\begin{eqnarray}
\nonumber \sum_{e \in E_{in}(v)}\! I_k(X_e)\!&=&\!\sum_{e \in
E_{out}(v)}\! I_k(X_e)\ \forall\ k,v \in {\cal V} - \{S, k\}
\\
\label{eqn:flowconst1}
 \sum_{e \in E_{out}(S)}\! I_k(X_e)\!&=&\!\sum_{e \in E_{in}(k)}\! I_k(X_e)\ = \ h\ \ \forall\ k
 \end{eqnarray}
\end{thm}
\begin{proof}
It is possible to decompose any multicast code into $h$ flows to
individual receivers \cite{ahlswede}. Consider the $X_e$'s and
$I_k(X_e)$'s corresponding to one such flow decomposition.
$I_k(X_e)$ is the amount of flow along edge $e$ in the flow
decomposition of receiver $k$. The equations in
(\ref{eqn:flowconst1}) claim that in the flow decomposition of
receiver $k$, the flow coming into any intermediate node is equal
to the flow coming out of the node, the flow coming out of the
source node is $h$, and, the flow into receiver $k$ is $h$. These
are well known properties of the flow decomposition
\cite{ahlswede}.
\end{proof}

It is convenient to define $X_{in}^{v}$ and $X_{out}^{v}$ for node
$v \in {\cal V}$ as $\sum_{e \in E_{in}(v)} X_e$ and $\sum_{e \in
E_{out}(v)} X_e$ respectively. To keep the notation brief we will
drop the superscript $v$ in discussions involving just one node.
Since $I$ is a linear function of $X$, the conditions in
(\ref{eqn:flowconst1}) reduce to
\begin{eqnarray}
\label{eqn:flowconst} \nonumber I_k(X_{in}^v) &=& I_k(X_{out}^v) \
\ \forall k,
\ \forall v \in {\cal V} - \{S, k\}\\
I_k(X_{out}^S) &=& I_k(X_{in}^{k})\ =\ h  \ \ \forall k
\end{eqnarray}
We will call the necessary conditions in
(\ref{eqn:flowconst}) as {\em flow constraints}. In Fig.
\ref{fig:flow} we can easily see that the edge and the flow
constraints are satisfied.

\subsection{Routing, Replicating and Network Coding}

Let us take a closer look at the different operations that occur in
a node in a multicast network. In Fig. \ref{fig:flow} we see that
there are three different operations that happen at a node. The
first and simplest operation is when a packet is routed to one of
the output edges. The second type of operation is replication in
which multiple copies of the packet are sent along different
edges. The third operation, network coding, refers to the case
when two or more packets are combined into one packet. We will see
that these three operations are sufficient to represent any
necessary processing being done at the node but before that we
need to understand what the different operations represent.


We first look at routing and replication. Each packet that comes
into a node has an associated set of receivers $Q \subset D$. The
packet has to eventually reach each node in $Q$. When it gets
routed onto one of the output edges, the packet on the output edge
still has to reach all nodes in $Q$. In terms of the information
flows to the various receivers, this corresponds to the case when
overlapping flows or a simple flow passes through a node and
continues unaffected.

When a packet gets replicated, then each copy of the packet on the
output edge has to reach nodes in $P_i$ a subset of $Q$. ($P_i$
has to be a subset of $Q$ since the packet has to reach only nodes
in $Q$.) Since the packet has to reach all nodes in $Q$ we have
$\cup P_i = Q$. Moreover, the same packet does not need to reach
the same destination along two different paths. Therefore the
$P_i's$ are disjoint $P_{i_1} \cap P_{i_2} =  \phi\ \forall i_1
\neq i_2$. In the flow decomposition replication corresponds to
the point where two or more overlapping flows diverge.

For example, consider the node 3 in Fig. \ref{fig:flow}. The
incoming packet has to be sent to both node 1 and node 2.
$X_{in}^3 = [0, 0, 1]$. The node replicates it and forwards it to
two edges. Along one of the edges that packet reaches node 1
$(X_{e(3,1)} = [1, 0, 0])$ and the packet sent on the other edge
is meant for node 2 $(X_{e(3,2)} = [0, 1, 0])$. At the output
$X_{out}^3 = [1, 1, 0]$.

This concept becomes clearer when we look at the relationship
between $X_{in}$ and $X_{out}$. We consider the case for two and
three destinations and then generalize the results. When there are
two destinations replication occurs only when a packet meant for
both destinations is replicated and sent on two different paths,
one path for each receiver node. $x_{in}(1,2)$ represents the
average number of packets coming in per unit time that need to go
to both $1$ and $2$. If $r (r \geq 0)$ of these packets are
duplicated and transmitted per unit time we have
\begin{eqnarray}
\label{tworeceivers} \nonumber
x_{out}(1) &=& x_{in}(1) + r \\
x_{out}(2) &=& x_{in}(2) + r \\
\nonumber x_{out}(1,2) &=& x_{in}(1,2) - r
\end{eqnarray}

Now consider the case with three destinations. Similar to the two
receiver case, a packet meant for two destinations can get
replicated to produce two packets for the two destinations. Let
$r_1$, $r_2$ and $r_3$ represent the amount of replication
corresponding to flows to receiver sets $\{\{1\},\{2\}\}$,
$\{\{1\},\{3\}\}$, and $\{\{2\},\{3\}\}$ respectively. When
packets meant for all three receivers replicate, they split the
flow in four possible ways  \{\{1\},\{2\},\{3\}\},
\{\{1\},\{2,3\}\}, \{\{2\},\{1,3\}\} and \{\{3\},\{1,2\}\}. Let
$r_4$, $r_5$, $r_6$ and $r_7$ represent the number of packets
replicated per unit time corresponding to the four cases.  The
relation between the $X_{in}$ and $X_{out}$ is therefore given by
\begin{eqnarray}
\nonumber
x_{out}(1) &=& x_{in}(1) + r_1 + r_2 + r_4 + r_5\\
\nonumber
x_{out}(2) &=& x_{in}(2) + r_1 + r_3 + r_4 + r_6\\
\nonumber
x_{out}(3) &=& x_{in}(3) + r_2 + r_3 + r_4 + r_7\\
x_{out}(1,2) &=& x_{in}(1,2)  - r_1 + r_7\\
\nonumber
x_{out}(1,3) &=& x_{in}(1,3)  - r_2 + r_6\\
\nonumber
x_{out}(2,3) &=& x_{in}(2,3)  - r_3 + r_5 \\
\nonumber x_{out}(1,2,3) &=& x_{in}(1,2,3) - r_4 -r_5 - r_6 - r_7
\end{eqnarray}

We note that each of the $r$'s are $\geq 0$. Moreover, if all the
$r$'s equal 0 then only routing is performed at a node. In the
general case we will have a routing variable $r_j$ associated with
every set $Q_j$ corresponding to flow for receivers in the set
$\cup_{Q \in Q_j}Q$ being replicated with each copy meant for a
set in $Q_j$.
We denote the set of routing variables $r_j$'s by $R$. The general
equation is
\begin{equation}
    x_{out}(P_i) = x_{in}(P_i) + \sum_{j:P_i \in Q_j}r_j - \sum_{j:\cup_{Q \in Q_j}Q = P_i}r_j
\label{eqn:routingconst}
\end{equation}

Any node that is restricted to routing/replicating has to satisfy
(\ref{eqn:routingconst}). We will call this constraint on $X_{in}$
and $X_{out}$ as {\em routing constraint}. Note that although we
call the variable $r_j$'s as routing variables they actually
correspond to replication. Also when we say a node is a routing
node we allow for replication at that node.

The third type of operation is network coding. This happens at
nodes where two or more flows merge. Similar to the routing
variables we define a set of network coding variables $N$ where
element $n_j$ represents the amount of flow meant for each set of
receivers $Q \in Q_j$ that merges to form one $n_j$ flow that has
to reach all receivers in the set $\cup_{Q \in Q_j}Q$. $n_j|Q_j|$
packets are network coded to form $n_j$ packets. It is easy to see
that for a network coding node the relationship between $X_{in}$
and $X_{out}$ has to be of the form
\begin{eqnarray}
    \nonumber
    x_{out}(P_i) = x_{in}(P_i) + \sum_{j:P_i \in Q_j}r_j - \sum_{j:\cup_{Q \in Q_j}Q=P_i}r_j \\ - \sum_{j:P_i \in Q_j}n_j + \sum_{j:\cup_{Q \in Q_j}Q = P_i}n_j
\end{eqnarray}
which reduces to
\begin{equation}
\label{eqn:netcodconst} x_{out}(P_i)= x_{in}(P_i) + \sum_{j:P_i
\in Q_j}(r_j-n_j) - \sum_{j:\cup_{Q \in Q_j}Q = P_i}(r_j-n_j)
\end{equation}

We note that it is sufficient to consider variables $r_j - n_j$
but we retain both for now. We will refer to the conditions in
(\ref{eqn:netcodconst}) as {\em node constraints}.

In the following theorem, for any pair of $X_{in}$ and $X_{out}$
that satisfy the flow constraints, we show that the operations at
the node can be decomposed into routing, replicating and network
coding operations and hence these operations are sufficient to
represent any processing done at the node.

\begin{thm} The relationship between $X_{in}$ and $X_{out}$ for
any valid operation at the node can be expressed in terms of
routing variables $R$ and network coding variables $N$ such that
each element of $R$ and $N$ is $\geq 0$ .
\end{thm}

\begin{proof}
We will give a particular solution satisfying all the conditions.
The main idea used in constructing the particular solution is that
all packets meant for more than one receiver can be replicated to
produce packets such that each packet is meant for one receiver.
They can then be suitably network coded to get the desired output
information flow vector.

Consider a set of receivers $P_i$ and corresponding set $Q(P_i)$
the set of all singleton subsets of $P_i$. For every set $P_i \in
{\cal P}$ containing two or more elements set $r_j = x_{in}(P_i)$
and $n_j = x_{out}(P_i)$ where $Q_j = Q(P_i)$. Set all other
routing and network coding variables to 0. We will show that this
solution satisfies the constraints in (\ref{eqn:netcodconst}). On
substituting for the routing and network coding variables that
have been set to 0, for all non singleton sets $P_i$ the
constraints in (\ref {eqn:netcodconst}) reduce to $x_{out}(P_i)=
x_{in}(P_i) - r_j + n_j$, $Q_j = Q(P_i)$ which is satisfied by
choice of $r_j$ and $n_j$. For singleton sets $P = \{k\}$ we have
\begin{eqnarray}
\nonumber
x_{out}(k) &=& x_{in}(k) + \sum_{j:\{k\} \in Q_j} (r_j-n_j)\\
\nonumber
&=& x_{in}(k) + \sum_{i:\{k\} \in Q_j = Q(P_i), |P_i|>1} (r_j-n_j)\\
\nonumber
   &=& x_{in}(k) + \sum_{i:k \in P_i, |P_i| > 1} x_{in}(P_i) - x_{out}(P_i)
\end{eqnarray}
which is exactly the flow constraint on information flow to the
receiver $k$ (Eq. \ref {eqn:flowconst}) and hence is satisfied.
\end{proof}

\begin{thm}
Given a network ${\cal G} = ({\cal V},{\cal E})$, flow vectors
$X_e$ for each edge $e \in {\cal E}$ and routing and network
coding variables $R^v$ and $N^v$ for each node $v \in {\cal V}$
such that the edge, flow and node constraints are satisfied, we
can construct a valid multicast code that performs routing and
network coding as specified by $R^v$ and $N^v$.
\label{lemma:construction}
\end{thm}
\begin{proof}
We prove the theorem by replacing each node in the network by a
network that has routing and network coding nodes corresponding to
the variables $R^v$ and $N^v$ such that there is no loss in the
multicast rate.

With every node $v \in {\cal V}$ associate a set of $(2^k -1)$
nodes where each new node, $v(P_i)$, corresponds to one set of
receivers $P_i \in {\cal P}$. For every set $P_i \in {\cal P}$
connect all the $x_{in}^v(P_i)$ incoming edges and the
$x_{out}^v(P_i)$ outgoing edges of node $v$ carrying data for
receivers in and only in set $P_i$ as input and output edges to
the node $v(P_i)$.

Corresponding to each non zero routing variable $r_j^v$ construct
$r_j^v$ nodes, each node having exactly one incoming edge coming
from node $v(\cup_{P_i \in Q_j} P_i)$ and $|Q_j|$ outgoing edges
that are connected as inputs to nodes in $\{v(P_i):P_i \in {\cal
Q_j}\}$. Corresponding to each non zero network coding variable
$n_j^v$ construct $n_j^v$ nodes with each node having one input
edge from every node in $\{v(P_i):P_i \in Q_j\}$ and one output
edge that is connected as input to node $v(\cup_{P_i \in Q_j}
P_i)$.

Now the number of incoming edges to node $v(P_i)$ is
$x_{in}^v(P_i) + \sum_{j:P_i \in Q_j}r_j^v + \sum_{j:\cup_{Q \in
Q_j}Q = P_i}n_j^v$ and the number of outgoing edges is
$x_{out}^v(P_i) + \sum_{j:P_i \in Q_j}n_j^v + \sum_{j:\cup_{Q \in
Q_j}Q = P_i}r_j^v$. From (\ref{eqn:netcodconst}) the number of
incoming edges is equal to the number of outgoing edges. Randomly
connect the set of input edges and the set of output edges of node
$v(P_i)$ in a one to one manner and delete node $v(P_i)$.

It is easy to see that this construction procedure replaces each
node by a network that maintains the same flows and hence there is
no loss in rate.
\end{proof}

In the construction procedure provided in the proof for Theorem
\ref{lemma:construction}, the network that replaces each node
could have cycles. These cycles are formed when a packet meant for
a set of receivers $P_i$ goes through a series of network coding
and routing operations to get back a packet meant for $P_i$
itself. Clearly the involvement of this packet in those operations
is unnecessary. All cycles correspond to unnecessary operations
and hence can be removed. We note that cycles within a node will
be absent in solutions that minimizes the number of network coding
operations. The construction procedure provided can be used along
with ideas of random network coding \cite{ho,ho2} to construct
multicast codes corresponding to the given information flow
vectors.


\section{Optimization}

\label{sec:optimize} Since any solution to the set of linear
equations specified by (\ref{eqn:edgeconst}),
(\ref{eqn:flowconst}) and (\ref{eqn:netcodconst}) corresponds to a
network coding solution, we can use the set of equations to obtain
a network coding solution in order to minimize a ``cost''
associated with the network code. The problem can be stated as
follows:
\begin{eqnarray}
\nonumber
\lefteqn{\mbox{minimize Cost}}\\
\nonumber \lefteqn{\mbox{subject to}}\\
\nonumber && x_e(P_i) \geq 0\ \forall\ P_i \in P,\ \forall\ e \in {\cal E},\\
\nonumber
&&r_j^v \geq 0, n_j^v \geq 0 \ \forall\ j\ \forall\ v \in {\cal V}\\
\nonumber \lefteqn{\mbox{\underline{Edge Constraints:} }} \\
\nonumber &&\sum_{P_i \in P} x_e(P_i) \leq C(e) \ \ \forall e \in
{\cal E} \\
\nonumber \lefteqn{\mbox{\underline{Node
Constraints:}}}\\\nonumber && x_{out}^v(P_i) =
x_{in}^v(P_i) + \sum_{j:P_i \in Q_j}\!(r_j^v-n_j^v) -\\
&&\nonumber\ \ \ \ \ \ \ \sum_{j:\cup_{Q \in Q_j}Q =
P_i}\!\!(r_j^v-n_j^v)\ \forall\ P_i \in {\cal P}\ \forall\
v \in {\cal V}\\
&&\nonumber I_k(X_{out}^S) = h\ \forall k\\
&&I_k(X_{in}^{k}) = h\ \forall\ k \label{eqn:constraints}
\end{eqnarray}
where
\begin{eqnarray}
\nonumber
    &&x_{in}^v(P_i) = \sum_{e\in E_{in}(v)}\!x_e(P_i),\ x_{out}^v(P_i) = \sum_{e\in
    E_{out}(v)}\!x_e(P_i)
    \\&&\nonumber \mbox{and } I_k(X_e) = \sum_{j : k \in P_j} x_e(P_j)
\end{eqnarray}
Note that we have dropped some of the flow constraints in
(\ref{eqn:flowconst}) as they are satisfied automatically if the
node constraints in (\ref{eqn:netcodconst}) are satisfied.

In the remainder of the section, we list a few natural cost
criteria.
\begin{enumerate}
\item {\bf Number of Network Coding nodes.} Since additional
coding capabilities are required at a node in order to perform
network coding, it is potentially of interest to minimize the
number of nodes performing network coding. Using Theorem
\ref{lemma:construction}, it follows that network coding needs to
be performed at a node $v$ only if $n_i^v
>0$ for some $i$. Since $n_i^v \geq 0$, this condition is
equivalent to $\sum_{i}n_i^v >0$. Thus, the number of nodes in the
network performing network coding is $\sum_{v \in {\cal V}}
I(\sum_{i}n_i^v >0)$, which we choose as the cost function.

However, note that for $n_i^v \geq 0$, the function $\sum_{v \in
{\cal V}} I(\sum_{i}n_i^v >0)$ is a concave function and the
problem becomes one of minimizing a concave function over a convex
set. This solution might admit local minima and standard convex
minimization techniques cannot be used to solve this problem. We
relax this problem and investigate minimizing the number of
network coding operations and minimizing the number of packets
involved in network coding in the following problems.

\item {\bf Number of network coding operations.} In this problem
we investigate minimizing the number of network coding operations
at a node $v$. From Theorem \ref{lemma:construction} it follows
that network coding operations (linear encoding of packets) need
to be performed corresponding to each $n_i^v$. Thus the number of
network coding operations at node $v$ is $\sum_i n_i^v$. We define
this quantity as the {\em amount} of network coding. Thus, the
cost function in this problem is given by $\sum_{v \in {\cal V}}
\sum_i n_i^v$.

\item {\bf Number of packets involved in network coding.} In this
problem we investigate minimizing the number of packets over which
network coding is performed at a node $v$. This is particularly
relevant in optical networks when a conversion from optical
signals to electrical signals is involved in order to encode the
packets. We conjecture that the cost function is given by $\sum_{v
\in {\cal V}}\sum_{i}\max(A_i^v,0)$ where $A_i^v = \sum_{j:P_i \in
Q_j}(n_j^v-\lambda_{i,j}^v) -$ $ \sum_{j:\cup_{Q \in Q_j}Q =
P_i}(n_j^v-\max_{i1}\lambda_{i1,j}^v)$. $\lambda_{i,j}$ represents
the number of packets meant for receivers $P_i$ that participate
in network coding and that are obtained by routing packets meant
for $\cup_{Q\in Q_j}Q$ ($P_i \in Q_j$). From the definition it
follows that $0 \leq \lambda_{i,j} \leq r_j$.


\item {\bf Minimum resource cost}. In the setup considered in
\cite{lun}, each edge $e$ is associated with a cost function
$f_e(z_e)$ when the data rate on $e$ is $z_e$. The net cost
associated with the network is then given by $\sum_e f_e(z_e)$.
This cost was minimized over the set of equations specified by
equations (1) and (2) in \cite{lun}. The same approach can be
applied in the setting where only certain nodes are allowed to
perform network coding. The restriction that a node $v$ can
perform only routing can be imposed by further constraining the
equations in (\ref{eqn:constraints}) by $n_i^v=0$ for all $i$.

\item {\bf Maximum rate}. The problem conidered here is one of
maximizing $h$ constrained to (\ref{eqn:constraints}) and
additionally the set of equations $n_i^v=0$ for all $i$ and nodes
$v$ which are restricted to routing.
\end{enumerate}

Note that the problems 2, 3, 4 (if the cost function $f_e()$ is
linear) and 5 are linear problems and can be solved by standard
linear programming approaches. It remains to be investigated if
the decentralized subgradient optimization suggested in \cite{lun}
 can be applied to these problems. To the end
of providing decentralized solutions to these linear problem, we
consider the approach suggested by \cite{lun} in which a linear
function $ax$ is approximated by a stricly convex function
$(ax)^{1+\alpha}$ where $\alpha >0$ is chosen small enough for a
valid approximation. This makes the problem a convex optimization
problem which can be solved in a decentralized manner by a
modified version of the primal-dual algorithm used in \cite{lun}.
We do not prove this due to lack of space. The main idea in the
proof is to show that the edge and node constraints involved are
local in the sense of involving variables of the neighbouring
edges or nodes and then follow the same steps as used in
\cite{lun}.

Problem 4 is a convex optimization problem if the function $f_e()$
is convex. If we further assume that the function $f_e()$ is
strictly convex, it follows that problem 4 admits a unique
solution. Further, it can be shown that the primal-dual algorithm
used in \cite{lun} can be modified to solve problem 4 in a
decentralized manner. Again we do not prove this due to lack of
space.

\section{Conclusion}
\label{sec:conclusion} In this paper, we presented a new
Information flow model to represent multicast flows. Using this
model we set up optimization problems and presented distributed
algorithms to minimize costs like number of packets undergoing
network coding and amount of network coding. We also showed that
this approach can be used to minimize network costs like link
usage when some nodes are restricted to routing.


\end{document}